\documentclass[journal]{IEEEtran}

\usepackage{multirow}
\usepackage{amsmath}
\usepackage{amssymb}
\usepackage{graphicx}
\usepackage{cite}
\usepackage{tabularx}
\usepackage{psfrag}

\newcommand{\eq}[1]{\begin{align*}#1\end{align*}}

\newcommand{\bb}{\boldsymbol{b}}
\newcommand{\be}{\boldsymbol{e}}
\newcommand{\bn}{\boldsymbol{n}}
\newcommand{\br}{\boldsymbol{r}}
\newcommand{\bu}{\boldsymbol{u}}
\newcommand{\bE}{\boldsymbol{E}}
\newcommand{\bF}{\boldsymbol{F}}
\newcommand{\bG}{\boldsymbol{G}}
\newcommand{\bH}{\boldsymbol{H}}
\newcommand{\R}{\mathbb{R}}
\newcommand{\Z}{\mathbb{Z}}

\newcommand{\sign}{\mathop{\mathrm{sign}}\nolimits}
\newcommand{\round}{\mathop{\mathrm{round}}\nolimits}
\newcommand{\roundc}{\mathop{\mathrm{roundc}}\nolimits}
\newcommand{\argmin}{\mathop{\mathrm{arg\,min}}\limits}

\newcommand{\algo}[1]{{\sffamily#1}}

\begin{document}

\title{Faster Projection in Sphere Decoding}

\author{Arash Ghasemmehdi and Erik Agrell}

\maketitle

\begin{abstract}
Most of the calculations in standard sphere decoders are redundant, in the sense that they either calculate quantities that are never used or calculate some quantities more than once. A new method, which is applicable to lattices as well as finite constellations, is proposed to avoid these redundant calculations while still returning the same result. Pseudocode is given to facilitate immediate implementation. Simulations show that the speed gain with the proposed method increases linearly with the lattice dimension. At dimension 60, the new algorithms avoid about 75 \% of all floating-point operations.
\end{abstract}


\IEEEpeerreviewmaketitle

\section{Introduction} \label{sec:1}

\IEEEPARstart{E}{very} lattice is represented with its \emph{generator matrix} $\bG$, whose entries are real numbers. Let $n$ and $m$ denote the number of rows and columns of $\bG$ respectively with $n \leq m$. The rows of $\bG$, which are $\bb_{1}, \ldots, \bb_{n}$, are called \emph{basis vectors} and are assumed to be linearly independent vectors in $\R^{m}$. The lattice of dimension $n$ is defined as the set of points
\begin{equation} \label{eq:0}
\Lambda(\bG,\Z) = \{u_{1}\bb_{1} + \ldots + u_{n}\bb_{n} \mid u_{i} \in \Z\}.
\end{equation}
This paper is about methods to find the \emph{closest point} in a lattice to a given vector $\br \in \R^{m}$, hereafter called \emph{received vector,} which requires minimization of the metric $\| \br-\bu\bG \|$ over all lattice points $\bu\bG$ with $\bu \in \Z^n$.

In 1981, Pohst \cite{ref:2} suggested a way of finding the closest point in lattices, which later on was complemented by Fincke and Pohst in \cite{ref:3}. The general method has later become known as \emph{sphere decoding.} The implementation details of the Fincke--Pohst (FP) enumeration method were first presented by Viterbo and Biglieri in \cite{ref:4}. In 1999, Viterbo and Boutros applied the FP enumeration method to maximum likelihood (ML) detection for finite constellations \cite{ref:5}. Later on, Agrell \emph{et al.}\ in \cite{ref:1} illustrated that the Schnorr-Euchner (SE) refinement \cite{ref:6} of the FP enumeration strategy improves the complexity of the sphere decoder algorithm.

During the last decade, a lot of work has been done to improve the efficiency of sphere decoder algorithms \cite{ref:13,ref:24,ref:26,ref:25,ref:27,ref:32}, due to the significant usage they have found in numerous types of applications. In communication theory, the closest point problem arises in ML detection for multiple-input multiple-output (MIMO) channels \cite{ref:13,ref:16,ref:18,ref:19}, ML sequence estimation \cite{ref:31}, quantization \cite{ref:28}, vector perturbation in multiuser communications \cite{ref:20}, and joint detection in direct-sequence multiple access system \cite{ref:21}.

The closest point search algorithms can be modified to find the ML point in finite constellations \cite{ref:13,ref:5}, which has an important application in MIMO channels. Assuming a system with $n$ transmit and $m$ receive antennas, the new set of points $\Lambda(\bG,\mathcal{U})$ is defined by replacing $\Z$ in \eqref{eq:0} with the finite range of integers
\begin{equation} \label{eq:17}
\mathcal{U}=\{U_\text{min}, U_\text{min}+1, \ldots, U_\text{max}\}.
\end{equation}
The transmit set can be mapped to an $L$-PAM constellation with $L=U_\text{max}-U_\text{min}+1$. The received vector after an additive white Gaussian noise (AWGN) channel with double-sided noise power spectral density $N_{0}/2$ is
\begin{equation} \label{eq:16}
\br=\bu\bG+\bn,
\end{equation}
where $\bu \in \mathcal{U}^{n}$, $\br \in \R^{m}$, $\bG \in \R^{n \times m}$, and $\bn \in \R^{m}$ is a vector of independent and identically distributed (i.i.d.) Gaussian noise with variance $N_{0}/2$. In this case, ML detection is equivalent to minimization of the metric $\| \br-\bu\bG \|$ over all possible points $\bu\bG$ with $\bu \in \mathcal{U}^{n}$. In MIMO systems where usually quadrature amplitude modulation (QAM) is used, the $L^{2}$-QAM signal constellation can be viewed as two real-valued $L$-PAM constellations with $\bu \in \mathcal{U}^{2n}$, $\br \in \R^{2m}$, $\bG \in \R^{2n \times 2m}$, and $\bn \in \R^{2m}$.

For both types of applications, lattices or finite constellations, the calculations can be implemented based on $\bG$, as in the original FP algorithm and its numerous refinements, notably \cite{ref:13,ref:31}, or based on $\bH = \bG^{-1}$ \cite{ref:1,ref:14}.

In this paper, we draw attention to a hitherto unnoticed problem with the standard algorithms. It is illustrated that the standard sphere decoder algorithms based on FP \cite{ref:3,ref:13} and SE \cite{ref:6,ref:13} enumeration strategies perform many excessive numerical operations. A method is proposed to avoid these unnecessary computations. However, the revision proposed is not related to choosing a more accurate upper bound on $\|\br-\bu\bG\|$ or scanning set of feasible point $\bu\bG$ in a different order. We believe that the SE strategy is the best way in this regard. Our modifications instead change how lattice vectors are recursively constructed from lower-dimensional lattices (for $\bG$-based implementations) or how the received vector $\br$ is recursively projected onto the basis vectors (for $\bH$-based implementations), which accounts for most of the floating point calculations in sphere decoding. With the proposed methods, not a single value would be calculated twice or remain without any use. Standalone implementations of the new (and old) algorithms are given in Fig.~\ref{fig:algorithms}.

\section{Closest Point Search Algorithms} \label{sec:2}

Without loss of generality, we assume that $\bG$ is a square lower-triangular matrix with positive diagonal elements \cite{ref:1}. Consequently, $\bH = \bG^{-1}$ is also square with positive diagonal elements. The decription of the sphere decoding principle in this section takes the $\bH$-based approach.

Every lattice can be divided into layers of lower-dimensional lattices. The diagonal elements of $\bH$ illustrate the distances between these layers, such that $1/H_{i,i}$ represents the distance between the $(i-1)$-dimensional layers in an $i$-dimensional layer. Thus, $1/H_{1,1}$ is the distance between the lattice points in a one-dimensional layer.

\begin{figure}
\psfrag{A}[cc][][0.75]{A}
\psfrag{B}[cc][][0.75]{B}
\psfrag{C}[cc][][0.75]{C}
\psfrag{D}[cc][][0.75]{D}
\psfrag{E}[cc][][0.75]{E}
\psfrag{r}[cc][][1]{$\br$}
\psfrag{rr}[cc][][1]{$\emph{~~\textbf{r}}_{n-1}$}
\psfrag{b}[cc][][1]{$\bb_{n}$}
\psfrag{y}[cc][][1]{$y_{n}$}
\psfrag{F}[cc][][1]{$\sqrt{C}$}
\psfrag{G}[cc][][1]{$1/H_{n,n}$}
\psfrag{H}[cc][][1]{$\hat{u}_{n}$}
\psfrag{I}[cc][][1]{$u_{n}$}
\centering
\includegraphics[width=8.0cm]{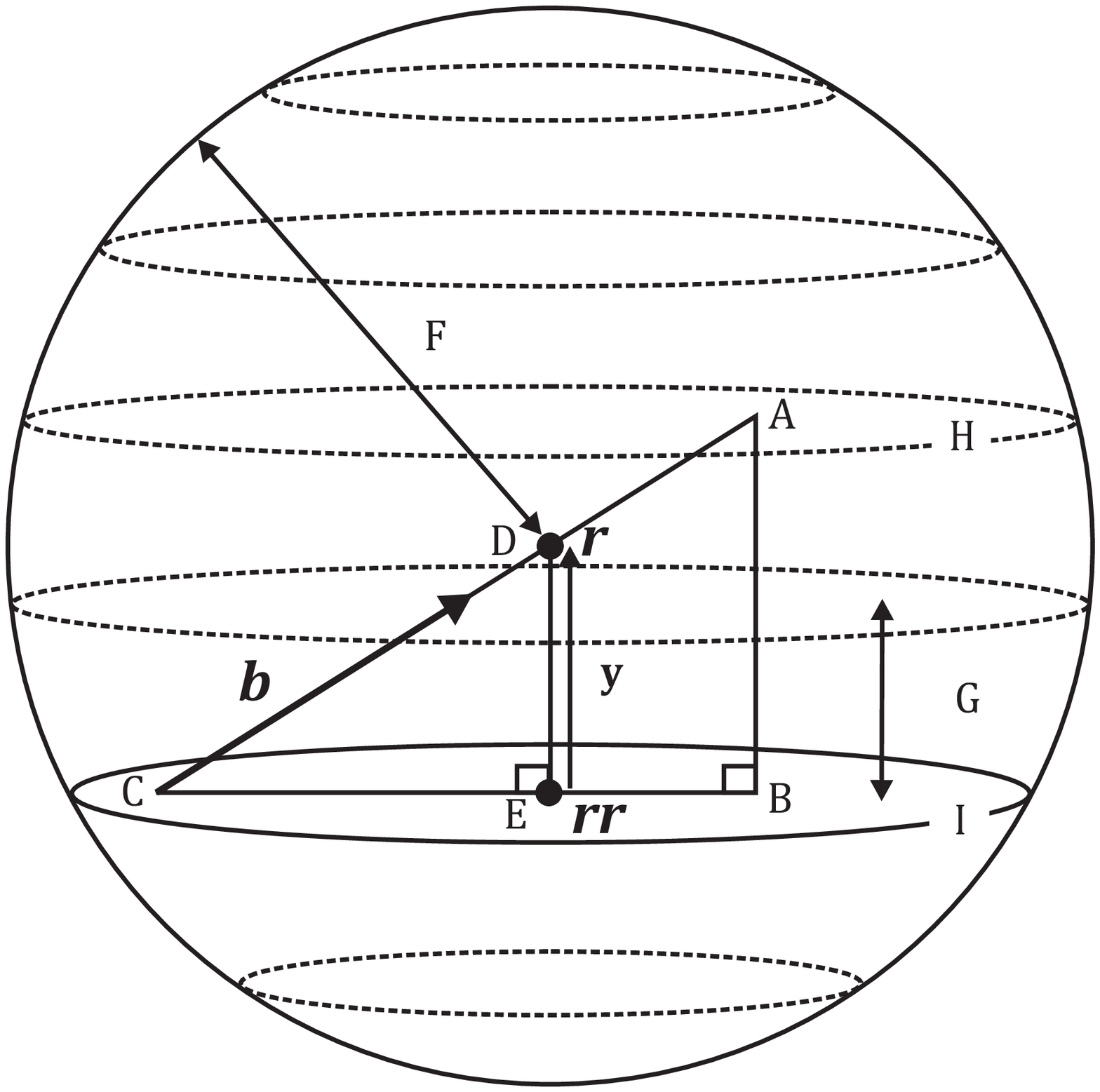}
\caption{Snapshot of an $n$-dimensional hypersphere, divided into a stack of $(n-1)$-dimensional hyperspheres (layers).} \label{fig:1}
\end{figure}

Fig.~\ref{fig:1} illustrates an $n$-dimensional hypersphere with radius $\sqrt{C}$ centered on a vector $\br$. All lattice points inside this hypersphere lie on $(n-1)$-dimensional layers, which are also hyperspheres. The basis vector $\bb_{n}$ is in the same direction as the hypotenuse of right triangles $\triangle ABC$ and $\triangle DEC$, while all the other basis vectors $\bb_{1},\ldots,\bb_{n-1}$ lie in the subspace spanned by one of these $(n-1)$-dimensional layers.

Starting from dimension $n$, the received vector $\br=(r_{1},r_{2},\ldots,r_{n}) \in \R^{n}$ is projected onto the lattice basis vectors $\bb_{1},\bb_{2},\ldots,\bb_{n}$. This is done by a simple matrix multiplication
$\be_{n}\bG=\br \Rightarrow \be_{n}=\br\bH$, where $\be_{n}=(E_{n,1},E_{n,2},\ldots,E_{n,n}) \in \R^{n}$. For known $C$ and $E_{n,n}$ the corresponding range for the integer component $u_{n}$ is \cite{ref:5}
\begin{equation} \label{eq:19}
\lceil -H_{n,n}\sqrt{C}+E_{n,n} \rceil \leq u_{n} \leq \lfloor H_{n,n}\sqrt{C}+E_{n,n} \rfloor,
\end{equation}
where $\lceil~\rceil$ and $\lfloor~\rfloor$ denote the round up and round down operations respectively, which is also intuitively conspicuous from Fig.~\ref{fig:1}.

For each $(n-1)$-dimensional layer $u_n$ that is to be examined, the orthogonal displacement $y_{n}$ from the received vector $\br$ to this layer is calculated, which is shown with line $\overline{DE}$ in Fig.~\ref{fig:1}. This displacement follows from the congruence of $\triangle ABC$ and $\triangle DEC$:
\begin{align}
\frac{(\hat{u}_{n}-u_{n})\frac{1}{H_{n,n}}}{y_{n}} & =\frac{(\hat{u}_{n}-u_{n})\|\bb_{n}\|}{(E_{n,n}-u_{n})\|\bb_{n}\|} \Rightarrow \nonumber \\
y_{n} & =\frac{E_{n,n}-u_{n}}{H_{n,n}}. \label{eq:18}
\end{align}

In order to calculate $E_{n-1,n-1}$, which will be used later on to calculate the range of $u_{n-1}$ \eqref{eq:21} and the displacement $y_{n-1}$ \eqref{eq:9}, the received vector $\br$ is first projected onto the examined $(n-1)$-dimensional layer \eqref{eq:23} and then to the lattice basis vectors \eqref{eq:24}. We use the notation $\br_{n-1}$ for the projected received vector $\br$, where $n-1$ denotes the dimension of the layer that the received vector is projected on.

Thanks to the lower-triangular representation, the orthogonal projection of $\br$ onto the $(n-1)$-dimensional layer currently being investigated affects only the last component of $\br$. Thus, it is sufficient to subtract $y_{n}$ from the $n$th element of $\br$ to obtain
\begin{equation} \label{eq:23}
\br_{n-1}=(r_{1},r_{2},\ldots,r_{n}-y_{n}).
\end{equation}
This positions $\br_{n-1}$ exactly on the perpendicular vertex of $\triangle DEC$. Projecting the vector $\br_{n-1}$ onto the lattice basis vectors can also be done by the multiplication
\begin{align}
\be_{n-1} & =\br_{n-1}\bH \label{eq:24} \\
& =\br\bH-(0,\ldots,0,y_{n})\bH \nonumber \\
& =\be_{n}-y_{n}(H_{n,1},\ldots,H_{n,n}), \label{eq:20}
\end{align}
where $\be_{n-1}=(E_{n-1,1},\ldots,E_{n-1,n-1},u_{n})$. 
The important element here is $E_{n-1,n-1}$, which is the value that should be multiplied to the lattice basis vector $\bb_{n-1}$ to create the projected vector $\br_{n-1}$. This element determines the corresponding range for $u_{n-1}$ \cite{ref:5}
\begin{align}
\lceil -H_{n-1,n-1} & \sqrt{C-\lambda_{n}}+E_{n-1,n-1} \rceil \leq u_{n-1} \nonumber \\
& \leq \lfloor H_{n-1,n-1}\sqrt{C-\lambda_{n}}+E_{n-1,n-1} \rfloor, \label{eq:21}
\end{align}
where $\lambda_{n}=y_{n}^{2}$ and $C-\lambda_{n}$ is the squared radius of the examined $(n-1)$-dimensional layer.

The sphere decoder is applied recursively to search this $(n-1)$-dimensional layer. Thereafter the next $u_{n}$ value in \eqref{eq:19} is generated and a new $(n-1)$-dimensional layer is searched. Generalizing, the closest point in an $i$-dimensional layer is found by dividing the layer into $(i-1)$-dimensional layers, searching each of these separately, and then proceeding to the next $i$-dimensional layer. We will refer to this process of decreasing and increasing $i$ as \emph{moving down and up the layers,} resp. We derive for $i=0,\ldots,n-1$
\begin{align}
\be_{i} & =\br_{i}\bH \label{eq:8} \\
& =\be_{n}-\sum_{j=i+1}^n y_{j}(H_{j,1},\ldots,H_{j,n}) \nonumber \\
& =\be_{i+1}-y_{i+1}(H_{i+1,1},\ldots,H_{i+1,n}), \label{eq:4}
\end{align}
where $\br_{i}$ is the received vector $\br$ projected onto an $i$-dimensional layer, and
\begin{equation} \label{eq:26}
\be_{i}=(E_{i,1},\ldots,E_{i,i},u_{i+1},\ldots,u_{n})
\end{equation}
gives the coefficients of $\br_{i}$ expressed as a linear combination of the lattice basis vectors. (In a zero-dimensional layer, which is a lattice point, $\br_{0} \in \Lambda(\bG,\Z)$ and $\be_{0}=\br_{0}\bH \in \Z^{n}$.)

Assuming an $i$-dimensional sphere similar to Fig.~\ref{fig:1}, the orthogonal displacement between the projected vector $\br_{i}$ and the examined $(i-1)$-dimensional layer is
\begin{equation} \label{eq:9}
y_{i}=\frac{E_{i,i}-u_{i}}{H_{i,i}}, ~~~~~ i=1,\ldots,n.
\end{equation}
Based on a lower-triangular form and the interpretation that $y_{i}$ only affects the $i$th component of $\br_{i}$, for $i=1,\ldots,n$
\begin{equation} \label{eq:7}
\br_{i-1}=(r_{1},\ldots,r_{i-1},r_{i}-y_{i},\ldots,r_{n}-y_{n}).
\end{equation}
Similarly, the bounds for every $i$-dimensional layer are
\begin{align}
\lambda_{i} & =y_{i}^2+y_{i+1}^2+\ldots+y_{n}^2,&  &i=1,\ldots,n, \label{eq:10} \end{align}
where $\lambda_{i}$ is the squared distance from the received vector $\br$ to the projected vector $\br_{i-1}$ and $C-\lambda_{i+1}$ is the squared radius of the examined $i$-dimensional layer. Hence, $\lambda_{1}$ denotes the Euclidean distance between the received vector $\br$ and a potential closest point $\br_{0}$. Finally, the range of $u_i$ for $i=1,\ldots,n-1$ is \cite{ref:5}
\begin{equation} \label{eq:22}
\lceil -H_{i,i}\sqrt{C-\lambda_{i+1}}+E_{i,i} \rceil \leq u_{i} \leq \lfloor H_{i,i}\sqrt{C-\lambda_{i+1}}+E_{i,i} \rfloor,
\end{equation}
where the \emph{projection value} $E_{i,i}$ is the value that should be multiplied with the lattice basis vector $\bb_{i}$ to create the projected vector $\br_{i}$.

\section{Avoiding Redundant Calculations} \label{sec:3}

In this section, we claim that most of the arithmetic operations in standard sphere decoders are redundant and we propose methods to avoid them, thus increasing the decoding speed. The redundant operations are of two types: for $\bH$-based implementations, numerous quantities are calculated which are never used, and for $\bG$-based implementations, some quantities are calculated more than once. In both cases, the source of the problem is the way the projection values are calculated.

\subsection{$\bH$-Based Decoding: Projection of The Received Vector} \label{sec:3.1}

Most of the numerical operations carried out in standard sphere decoders based on $\bH$ are related to the projection of the received vector $\br$, or its lower-dimensional counterpart, onto the lattice basis vectors as in \eqref{eq:8}--\eqref{eq:26}. Defining a matrix $\bE$ whose rows are $\be_1,\ldots,\be_n$, it follows from \eqref{eq:4} that all elements of this matrix are updated from the elements immediately below. However, the only values that are required in the sphere decoder algorithms are the diagonal elements $E_{i,i}$, used in \eqref{eq:9} and \eqref{eq:22}. Thus, the elements located above the diagonal of $\bE$ are not required to be calculated. They correspond to $u$ values that have already been calculated in previous stages of the algorithm, see \eqref{eq:26}.

The sphere decoder proposed in \cite{ref:1} always updates the first $i$ elements of $\be_{i}$ simultaneously. For instance, if we are in an $i$-dimensional layer after computing $E_{i,i}$, we update $E_{i,j}$ for all $j=1,\ldots,i-1$.
These values may be used later to update $E_{j,j}$ for some $j<i$ after moving down the layers. But why should one project the entire vector $\br_{i}$ to the lattice basis vectors, and calculate the $E_{i,j}$ for all $j=1,\ldots,i-1$, when they are not supposed to be used at that stage of the algorithm, and possibly not at all? The answer to this question inspires an intelligent algorithm to manage the projection of $\br$ and updating the $E_{j,i}$ values, based on following criteria: \newline
$\bullet$ As explained in Sec.~\ref{sec:3.1}, we are just interested in elements located in the lower triangular form of $\bE$. \newline
$\bullet$ The last row of $\bE$, $\be_{n}$, is just calculated once since there exists just a single $n$-dimensional layer.  \newline
$\bullet$ According to \eqref{eq:4} and \eqref{eq:26}, updating an element $E_{j,i}$ (with $i \leq j < n$) requires knowledge of both $E_{j+1,i}$ and $y_{j+1}$.  \newline
$\bullet$ Unlike the row-wise updating method in \cite{ref:1}, we propose updating $\bE$ column-wise, i.e., updating $E_{j,i},E_{j-1,i},\ldots,E_{i+1,i}$ for a suitable value of $j$ before calculating the desired value of $E_{i,i}$.  \newline
$\bullet$ If we move to an $i$-dimensional layer, the first $i+1$ elements of $\be_{i}$ and of other $\be$ vectors above that row will be affected, since we are projecting the received vector $\br$ to this new $i$-dimensional layer. However, the elements below $\be_{i}$ will remain unaffected.  \newline
$\bullet$ Our main target at each stage, when we are moving towards the lower-dimensional layers, is just to update the $E_{i,i}$ values. The other $E_{j,i}$ values for $j>i$ will be updated if and only if they are needed to calculate the $E_{i,i}$ values.  \newline
$\bullet$ The algorithm should track of the movement down and up the layers in order to avoid the recalculation of values that remain unchanged, see Sec.~\ref{sec:3.3}.

In Sec.~\ref{sec:5}, we demonstrate by simulations how the complexity of sphere decoder algorithms, for both lattices and finite constellations, is reduced due to the method outlined above for projecting of the received vector $\br$.

\subsection{$\bG$-Based Decoding: Updating The Projection Values} \label{sec:3.2}

Also in the $\bG$-based implementations, the time-consuming step is to calculate the projection values, which we denote with $E_{i,i}$ in $\bH$-based implementations, as discussed in Sec.~\ref{sec:3.1}, and $p_i$ in $\bG$-based implementations.

According to \cite{ref:13}, which uses the same recursions as \cite{ref:5}, the projection value is calculated as $p_{i}=(r_{i}-f_{i})/G_{i,i}$, where $f_{n}=0$ and
\begin{equation} \label{eq:31}
f_{i}=\sum_{k=i+1}^n u_{k}G_{k,i}, ~~~~~ i=1,\ldots,n-1.
\end{equation}
Moving further down the layers in order to calculate the $p_{j}$ projection value for $j<i$, one can notice that part of the sum in \eqref{eq:31} is already calculated and does not need to be recalculated if stored in memory. Hence, we define $F_{j,i}=\sum_{k=j+1}^n u_{k}G_{k,i}$ for $1 \le i \le j < n$ and $F_{n,i} = 0$ for $1 \le i \le n$. As a result, we can calculate
\begin{equation} \label{eq:32}
F_{j-1,i}=F_{j,i}+u_{j}G_{j,i}
\end{equation}
for $1 \le i < j \le n$ and $p_{i}=(r_{i}-F_{i,i})/G_{i,i}$ for $1 \le i \le n$, which requires fewer operations than \eqref{eq:31}.

We collect the elements $F_{j,i}$ in a lower-triangular matrix $\bF$, which is completely irrelevant to the matrix $\bE$ discussed in Sec.~\ref{sec:2} and \ref{sec:3.1}. However, the optimized projection method proposed in Sec.~\ref{sec:3.1} to update the $E_{j,i}$ values, with some minor modifications, can be similarly applied herein to update the $F_{j,i}$ values. The changes are as follows: \newline
$\bullet$ The last row of $\bF$ is the zero vector.  \newline
$\bullet$ According to \eqref{eq:32}, updating $F_{j,i}$ requires knowledge of both $F_{j+1,i}$ and $u_{j+1}$ values. \newline
$\bullet$ If we move to an $i$-dimensional layer, all the elements of the $i$th row of $\bF$, and of other rows above that row, will be affected, since we are investigating a new $i$-dimensional layer. However, the elements below that row will remain unaffected.  \newline
$\bullet$ Our main target at each stage, when we are moving towards the lower-dimensional layers, is just to update the $F_{i,i}$ values. The other $F_{j,i}$ values for $j>i$ will be updated if and only if they are needed to calculate the $F_{i,i}$ values.  \newline
$\bullet$ Similarly to Sec.~\ref{sec:3.1}, we should keep track of the movement up and down the layers.

Based on the preceding criteria, one can avoid starting from the $n$th layer and updating all $F_{n-1,i},F_{n-2,i},\ldots,F_{i+1,i}$ elements located in the $i$th column of $\bF$ before calculating the objective $F_{i,i}$ value. While this significantly reduces the complexity of the algorithm, the memory write operations are increased.

\subsection{The Proposed Algorithm} \label{sec:3.3}

\begin{figure}
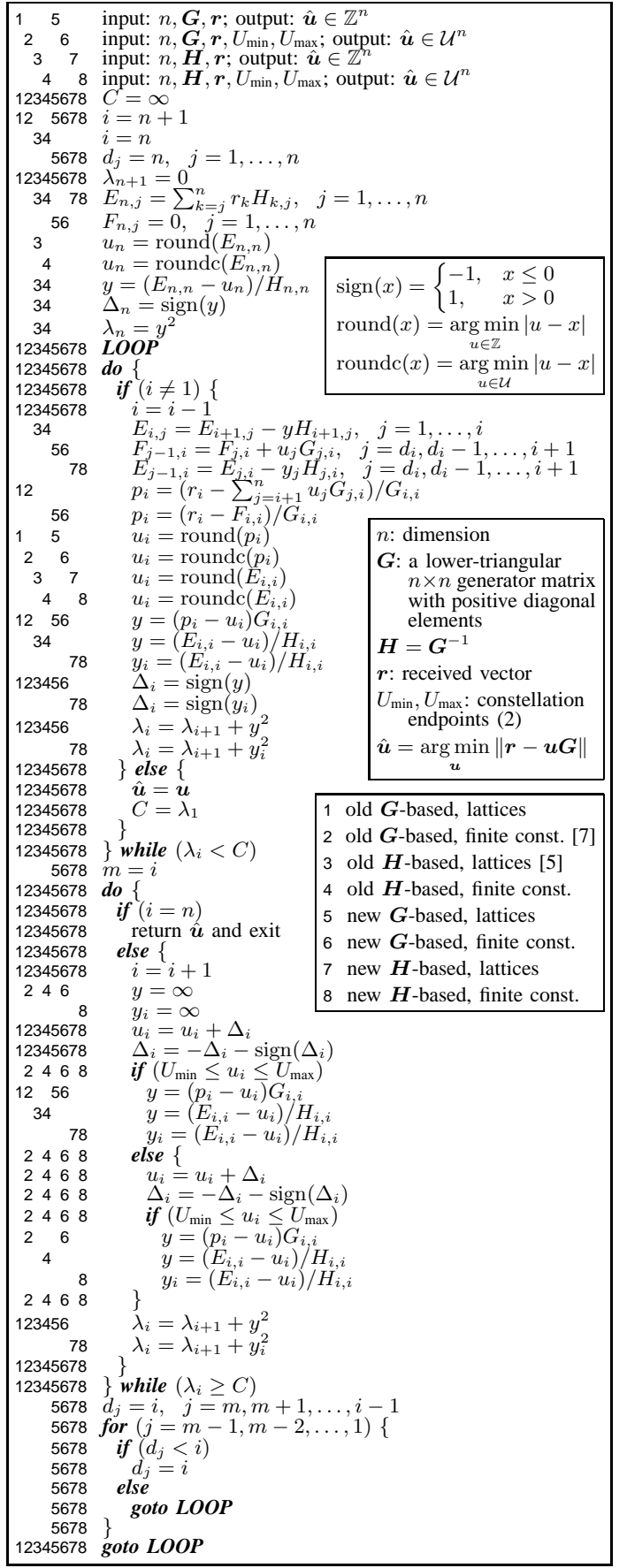

\newcommand{\e}[1]{\algo{\scriptsize#1}~}
\newcommand{\s}{\hspace*{.555em}} 
\fbox{%
{\renewcommand{\baselinestretch}{0.85}%
\parbox{\columnwidth}{%
\small%
\e{1\s\s\s5\s\s\s} input: $n,\bG,\br$; output: $\hat{\bu}  \in \Z^n$\\
\e{\s2\s\s\s6\s\s} input: $n,\bG,\br,U_\text{min},U_\text{max}$; output: $\hat{\bu}  \in \mathcal{U}^n$\\
\e{\s\s3\s\s\s7\s} input: $n,\bH,\br$; output: $\hat{\bu}  \in \Z^n$\\
\e{\s\s\s4\s\s\s8} input: $n,\bH,\br,U_\text{min},U_\text{max}$; output: $\hat{\bu}  \in \mathcal{U}^n$\\
\e{12345678} $C=\infty$\\
\e{12\s\s5678} $i=n+1$\\
\e{\s\s34\s\s\s\s} $i=n$\\
\e{\s\s\s\s5678} $d_{j}=n, ~~ j=1,\ldots,n$\\
\e{12345678} $\lambda_{n+1}=0$\\
\e{\s\s34\s\s78} $E_{n,j} = \sum_{k=j}^n r_k H_{k,j}, ~~ j=1,\ldots,n$\\
\e{\s\s\s\s56\s\s} $F_{n,j} = 0, ~~ j=1,\ldots,n$\\
\e{\s\s3\s\s\s\s\s} $u_{n}=\round(E_{n,n})$\\
\e{\s\s\s4\s\s\s\s} $u_{n}=\roundc(E_{n,n})$\\
\e{\s\s34\s\s\s\s} $y=(E_{n,n}-u_{n})/H_{n,n}$\\
\e{\s\s34\s\s\s\s} $\Delta_{n}=\sign(y)$\\
\e{\s\s34\s\s\s\s} $\lambda_{n}=y^2$\\
\e{12345678} $\emph{\textbf{LOOP}}$\\
\e{12345678} $\emph{\textbf{do}}~ \{$\\
\e{12345678} $~~\emph{\textbf{if}}~(i\neq1)~\{$\\
\e{12345678} $~~~~i=i-1$\\
\e{\s\s34\s\s\s\s} $~~~~E_{i,j}=E_{i+1,j}-yH_{i+1,j}, ~~ j=1,\ldots,i$\\
\e{\s\s\s\s56\s\s} $~~~~F_{j-1,i}=F_{j,i}+u_{j}G_{j,i}, ~~ j=d_{i},d_{i}-1,\ldots,i+1$\\
\e{\s\s\s\s\s\s78} $~~~~E_{j-1,i}=E_{j,i}-y_{j}H_{j,i}, ~~j=d_{i},d_{i}-1,\ldots,i+1$\\
\e{12\s\s\s\s\s\s} $~~~~p_{i}=(r_{i}-\sum_{j=i+1}^n u_j G_{j,i})/G_{i,i}$\\
\e{\s\s\s\s56\s\s} $~~~~p_{i}=(r_{i}-F_{i,i})/G_{i,i}$\\
\e{1\s\s\s5\s\s\s} $~~~~u_{i}=\round(p_{i})$\\
\e{\s2\s\s\s6\s\s} $~~~~u_{i}=\roundc(p_{i})$\\
\e{\s\s3\s\s\s7\s} $~~~~u_{i}=\round(E_{i,i})$\\
\e{\s\s\s4\s\s\s8} $~~~~u_{i}=\roundc(E_{i,i})$\\
\e{12\s\s56\s\s} $~~~~y=(p_{i}-u_{i})G_{i,i}$\\
\e{\s\s34\s\s\s\s} $~~~~y=(E_{i,i}-u_{i})/H_{i,i}$\\
\e{\s\s\s\s\s\s78} $~~~~y_{i}=(E_{i,i}-u_{i})/H_{i,i}$\\
\e{123456\s\s} $~~~~\Delta_{i}=\sign(y)$\\
\e{\s\s\s\s\s\s78} $~~~~\Delta_{i}=\sign(y_{i})$\\
\e{123456\s\s} $~~~~\lambda_{i}=\lambda_{i+1}+y^2$\\
\e{\s\s\s\s\s\s78} $~~~~\lambda_{i}=\lambda_{i+1}+y_{i}^2$\\
\e{12345678} $~~\}~\emph{\textbf{else}}~\{$\\
\e{12345678} $~~~~\hat{\bu}=\bu$\\
\e{12345678} $~~~~C=\lambda_{1}$\\
\e{12345678} $~~\}$\\
\e{12345678} $\}~\emph{\textbf{while}}~(\lambda_{i}<C)$\\
\e{\s\s\s\s5678} $m=i$\\
\e{12345678} $\emph{\textbf{do}}~\{$\\
\e{12345678} $~~\emph{\textbf{if}}~(i=n)$\\
\e{12345678} $~~~~\text{return }\hat{\bu} \text{ and exit}$\\
\e{12345678} $~~\emph{\textbf{else}}~\{$\\
\e{12345678} $~~~~i=i+1$\\
\e{\s2\s4\s6\s\s} $~~~~y=\infty$\\
\e{\s\s\s\s\s\s\s8} $~~~~y_{i}=\infty$\\
\e{12345678} $~~~~u_{i}=u_{i}+\Delta_{i}$\\
\e{12345678} $~~~~\Delta_{i}=-\Delta_{i}-\sign(\Delta_{i})$\\
\e{\s2\s4\s6\s8} $~~~~\emph{\textbf{if}}~(U_\text{min} \leq u_{i} \leq U_\text{max})$\\
\e{12\s\s56\s\s} $~~~~~~y=(p_{i}-u_{i})G_{i,i}$\\
\e{\s\s34\s\s\s\s} $~~~~~~y=(E_{i,i}-u_{i})/H_{i,i}$\\
\e{\s\s\s\s\s\s78} $~~~~~~y_{i}=(E_{i,i}-u_{i})/H_{i,i}$\\
\e{\s2\s4\s6\s8} $~~~~\emph{\textbf{else}}$ \{\\
\e{\s2\s4\s6\s8} $~~~~~~u_{i}=u_{i}+\Delta_{i}$\\
\e{\s2\s4\s6\s8} $~~~~~~\Delta_{i}=-\Delta_{i}-\sign(\Delta_{i})$\\
\e{\s2\s4\s6\s8} $~~~~~~\emph{\textbf{if}}~(U_\text{min} \leq u_{i} \leq U_\text{max})$\\
\e{\s2\s\s\s6\s\s} $~~~~~~~~y=(p_{i}-u_{i})G_{i,i}$\\
\e{\s\s\s4\s\s\s\s} $~~~~~~~~y=(E_{i,i}-u_{i})/H_{i,i}$\\
\e{\s\s\s\s\s\s\s8} $~~~~~~~~y_{i}=(E_{i,i}-u_{i})/H_{i,i}$\\
\e{\s2\s4\s6\s8} $~~~~\}$\\
\e{123456\s\s} $~~~~\lambda_{i}=\lambda_{i+1}+y^2$\\
\e{\s\s\s\s\s\s78} $~~~~\lambda_{i}=\lambda_{i+1}+y_{i}^2$\\
\e{12345678} $~~\}$\\
\e{12345678} $\}~\emph{\textbf{while}}~(\lambda_{i} \geq C)$\\
\e{\s\s\s\s5678} $d_{j}=i, ~~ j=m,m+1,\ldots,i-1$\\
\e{\s\s\s\s5678} $\emph{\textbf{for}}~(j=m-1,m-2,\ldots,1)~\{$\\
\e{\s\s\s\s5678} $~~\emph{\textbf{if}}~(d_{j}<i)$\\
\e{\s\s\s\s5678} $~~~~d_{j}=i$\\
\e{\s\s\s\s5678} $~~\emph{\textbf{else}}$\\
\e{\s\s\s\s5678} $~~~~\emph{\textbf{goto}} ~ \emph{\textbf{LOOP}}$
  \hfill\raisebox{132ex}[0ex][0ex]{\fbox{\parbox[t]{0cm}{
  \vspace*{-2ex}\eq{
  &\sign(x) = \begin{cases}-1, & x\le0 \\ 1, & x>0 \end{cases}\\[-.5ex]
  &\round(x) = \argmin_{u\in\Z}|u-x|\\[-.5ex]
  &\roundc(x) = \argmin_{u\in\mathcal{U}}|u-x|
  }\vspace*{-2ex}
  }}}\\
\e{\s\s\s\s5678} $\}$
  \hfill\raisebox{105ex}[0ex][0ex]{\fbox{\parbox[t]{3.3cm}{
  $n$: dimension \\[.7ex]
  $\bG$: a lower-triangular\\\hspace*{1.5em}$n\times n$ generator matrix\\\hspace*{1.5em}with positive diagonal\\\hspace*{1.5em}elements \\[.7ex]
  $\bH = \bG^{-1}$\\[.7ex]
  $\br$: received vector \\[.7ex]
  $U_\text{min},U_\text{max}$: constellation\\\hspace*{1.5em}endpoints \eqref{eq:17}\\[.7ex]
  $\hat{\bu} = \argmin_{\bu} \| \br-\bu\bG \|$
  }}}\\
\e{12345678} $\emph{\textbf{goto}} ~ \emph{\textbf{LOOP}}$
  \hfill\raisebox{78ex}[0ex][0ex]{\fbox{\parbox[t]{4.1cm}{
  \e{1}\hspace{0.9mm}old $\bG$-based, lattices \\[.7ex]
  \e{2}~old $\bG$-based, finite const.~\cite{ref:13}\\[.7ex]
  \e{3}~old $\bH$-based, lattices \cite{ref:1} \\[.7ex]
  \e{4}~old $\bH$-based, finite const. \\[.7ex]
  \e{5}~new $\bG$-based, lattices \\[.7ex]
  \e{6}~new $\bG$-based, finite const. \\[.7ex]
  \e{7}~new $\bH$-based, lattices \\[.7ex]
  \e{8}~new $\bH$-based, finite const.
  }}}
}}}
\caption{Eight algorithms in one figure. To implement a certain algorithm, use only the lines labeled with the algorithm's digit \algo{1,\ldots,8}.}
\label{fig:algorithms}
\end{figure}

Standalone representations of the old and new algorithms, $\bG$-based and $\bH$-based versions, for lattices and finite constellations, are given in Fig.~\ref{fig:algorithms}, all based on the SE enumeration strategy. The specifications are intended to be sufficiently detailed to allow a straightforward implementation, even without knowledge of the underlying theory.

As starting points, we use the $\bG$-based algorithm called ``Algorithm II'' in \cite{ref:13}, labeled with \algo{2} in Fig.~\ref{fig:algorithms}, and the $\bH$-based algorithm ``Decode'' in \cite{ref:1}, here labeled with \algo{3}. The loops have been restructured for consistency between the algorithms, but the calculations in Fig.~\ref{fig:algorithms} are exactly the same as in \cite{ref:13,ref:1}. Indeed, all algorithms for lattice decoding (algorithms \algo{1}, \algo{3}, \algo{5}, and \algo{7}) visit the same layers $u_i$, in the same order, and return the same result $\hat{\bu}$, although they calculate different intermediate quantities. A similar note holds for decoding finite constellations (algorithms \algo{2}, \algo{4}, \algo{6}, and \algo{8}).

After the initialization, the algorithms are divided into three parts. In the first part, we move down the layers (decrease $i$), as long as the squared Euclidean distance $\lambda_{i}$ \eqref{eq:10} between the received vector $\br$ and the projected vector $\br_{i-1}$ \eqref{eq:7} is less than the squared Euclidean distance $C$ between the received vector $\br$ and the closest lattice point detected so far. In the second part, we move up in the hierarchy of layers (increase $i$) as long as $\lambda_{i} \ge C$. Moreover, before leaving each of these parts, we store the minimum and maximum level $i$ that has been visited. These values are used in the last part of the algorithm, which only belongs to the new algorithms.

The method to manage the recursive projection of $\be_{i}$ \eqref{eq:8} or the calculation of $F_{j,i}$ is proposed in the last part. The value of $d_j$ for $j=1,\ldots,n$ denotes the starting point for the recursions in \eqref{eq:4} and \eqref{eq:32} in order to update the objective $E_{i,i}$ or $F_{i,i}$ values. For instance, $d_{i}=k$ indicates that in order to update $E_{i,i}$, we should start the projection from $k$th layer, where $k>i$, and calculate $E_{j,i}$ for $j=k-1,k-2,\ldots,i$.

Due to the well-documented performance gain that the SE enumeration strategy brings to sphere decoders, we apply herein the proposed refinement only to the SE strategy. However, the same refinement can be applied to the original FP enumeration strategy. It is also applicable to most, or all, of the numerous sphere decoder variants, optimal as well as suboptimal, that have been developed in the last decade.

\section{Simulation Results} \label{sec:5}

Herein, we evaluate the effectiveness of the proposed smart vector projection technique on the sphere decoder algorithms based on SE enumeration strategy, for both lattices and finite constellations. All eight algorithms are implemented according to the pseudocode presented in Fig.~\ref{fig:algorithms}.

We base our performance comparison measure on counting the number of floating point operations (flops) and integer operations (intops) that each algorithm carries out to reach the closest lattice point. Both types of operations include addition, subtraction, multiplication, division, and comparison, but not $\emph{\textbf{for}}$ loop counters, whose role differs between programming languages. The $\emph{\textbf{round}}$ operation is counted as a single floating point operation, and $\emph{\textbf{roundc}}$ in Sec.~\ref{sec:5.2} is counted as one floating point operation for 2-PAM and two for 4-PAM.

To compare the complexity of two algorithms, typically an old and a new one, we generate $M$ random generator matrices $\bG_{1},\ldots,\bG_{M}$, and for each $\bG_{j}$ we generate $N$ random received vectors $\br_{j,1},\ldots,\br_{j,N}$. The same vectors are decoded using both algorithms and the number of operations $\mathit{ops}(\br_{j,i},\bG_{j})$ is counted, which could be either flops or intops.
The average gain with the new algorithm is reported as
\begin{equation} \label{eq:13}
\mathit{gain}=\frac{1}{M} \sum_{j=1}^M \frac{\sum_{i=1}^N \mathit{ops}_\text{old}(\br_{j,i},\bG_{j})}{\sum_{i=1}^N \mathit{ops}_\text{new}(\br_{j,i},\bG_{j})}
.\end{equation}

\subsection{Lattices} \label{sec:5.1}

We generate the lattice generator matrices with random numbers, drawn from i.i.d.\ zero-mean, unit-variance Gaussian distributions. The random input vectors are generated uniformly inside a Voronoi region according to \cite{ref:12}. Our simulation results are based on averaging over $M=100$ different generator matrices. The number of input vectors $N$ depends on the dimension $n$ of the lattices. Fewer input vectors are examined in high dimensions, to the extent that we ensure that the plotted curves are reasonably smooth.

\begin{figure}[hb]
\psfrag{Dimension}[cc][][0.8]{Dimension}
\psfrag{Flops}[][][0.8]{Flops}
\psfrag{Title}[cc][][0.8]{}
\psfrag{G Based New}[cc][][0.7]{\hspace{9.1mm}$\bG$-based new}
\psfrag{G Based Old}[cc][][0.7]{\hspace{8.8mm}$\bG$-based old}
\psfrag{H Based New}[cc][][0.7]{\hspace{8.5mm}$\bH$-based new}
\psfrag{H Based Old}[cc][][0.7]{\hspace{8.3mm}$\bH$-based old}
\psfrag{0}[cc][][0.8]{\vspace{27.9mm}$0$}
\psfrag{5}[cc][][0.8]{$5$}
\psfrag{10}[cc][][0.8]{$10$}
\psfrag{15}[cc][][0.8]{$15$}
\psfrag{20}[cc][][0.8]{$20$}
\psfrag{25}[cc][][0.8]{$25$}
\psfrag{30}[cc][][0.8]{$30$}
\psfrag{35}[cc][][0.8]{$35$}
\psfrag{40}[cc][][0.8]{$40$}
\psfrag{45}[cc][][0.8]{$45$}
\psfrag{50}[cc][][0.8]{$50$}
\psfrag{55}[cc][][0.8]{$55$}
\psfrag{60}[cc][][0.8]{$60$}
\psfrag{A}[cc][][0.8]{$10^{2}$~~~~~}
\psfrag{B}[cc][][0.8]{$10^4$~~~~~}
\psfrag{C}[cc][][0.8]{$10^6$~~~~~}
\psfrag{D}[cc][][0.8]{$10^8$~~~~~}
\psfrag{E}[cc][][0.8]{$10^{10}$~~~~}
\psfrag{F}[cc][][0.8]{$10^{12}$~~~~}
\psfrag{G}[cc][][0.8]{\hspace{1.0mm}$10^{14}$~~~~}
\centering
{\includegraphics[height=5.5cm, width=\columnwidth,
trim=10mm 0mm 35mm 15mm, clip]{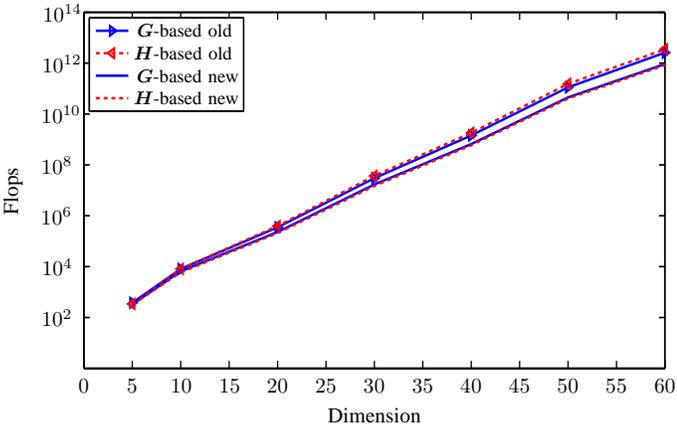}}
\caption{The average number of flops needed to decode a vector with the old and new versions of $\bG$- and $\bH$-based lattice decoding algorithms, without reduction.} \label{fig:flops}
\end{figure}

Fig.~\ref{fig:flops} compares the number of flops for the standard $\bG$- and $\bH$-based algorithms (algorithms \algo{1} and \algo{3} in Fig.~\ref{fig:algorithms}) with the new algorithms proposed in this paper (algorithms \algo{5} and \algo{7}). It can be seen that the $\bG$- and $\bH$-based implementations have about the same complexity, but both can be significantly improved.

\begin{figure}
\psfrag{Dimension}[cc][][0.8]{Dimension}
\psfrag{Ratio}[][][0.8]{Gain}
\psfrag{Operations}[cc][][0.8]{}
\psfrag{G Based Without Reduction}[cc][][0.7]{\hspace{26mm}$\bG$-based flops  without reduction}
\psfrag{G Based With LLL Reduction}[cc][][0.7]{\hspace{28.1mm}$\bG$-based flops with LLL reduction}
\psfrag{H Based Without Reduction}[cc][][0.7]{\hspace{25.8mm}$\bH$-based flops  without reduction}
\psfrag{H Based With LLL Reduction}[cc][][0.7]{\hspace{27.9mm}$\bH$-based flops with LLL reduction}
\psfrag{G H Based Without Reduction}[cc][][0.7]{\hspace{32.5mm}$\bG$\&$\bH$-based intops without reduction}
\psfrag{G H Based With LLL Reduction}[cc][][0.7]{\hspace{34.6mm}$\bG$\&$\bH$-based intops with LLL reduction}
\psfrag{0}[cc][][0.8]{0}
\psfrag{5}[cc][][0.8]{5}
\psfrag{10}[cc][][0.8]{10}
\psfrag{15}[cc][][0.8]{15}
\psfrag{20}[cc][][0.8]{20}
\psfrag{25}[cc][][0.8]{25}
\psfrag{30}[cc][][0.8]{30}
\psfrag{35}[cc][][0.8]{35}
\psfrag{40}[cc][][0.8]{40}
\psfrag{45}[cc][][0.8]{45}
\psfrag{50}[cc][][0.8]{50}
\psfrag{55}[cc][][0.8]{55}
\psfrag{60}[cc][][0.8]{60}
\psfrag{0.5}[cc][][0.8]{0.5~}
\psfrag{1}[cc][][0.8]{1~}
\psfrag{1.5}[cc][][0.8]{1.5~}
\psfrag{2}[cc][][0.8]{2~}
\psfrag{2.5}[cc][][0.8]{2.5~}
\psfrag{3}[cc][][0.8]{3~}
\psfrag{3.5}[cc][][0.8]{3.5~}
\psfrag{4}[cc][][0.8]{4~}
\psfrag{4.5}[cc][][0.8]{4.5~}
\centering
{\includegraphics[height=5.5cm, width=\columnwidth,
trim=10mm 0mm 35mm 15mm, clip]{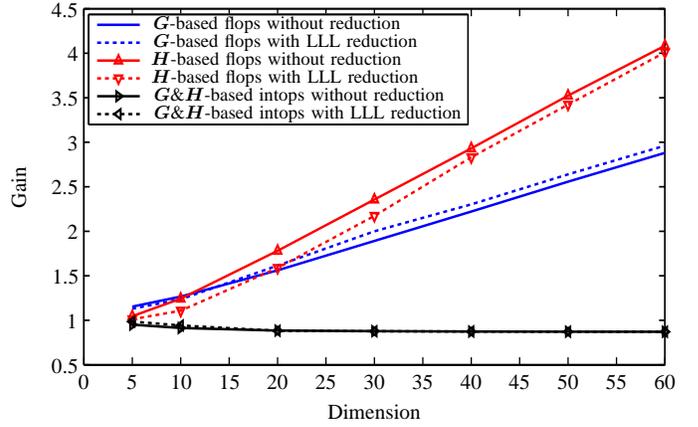}}
\caption{Complexity gain with the new lattice decoding algorithms, with and without reduction.} \label{fig:2}
\end{figure}

The gain \eqref{eq:13} is shown in Fig.~\ref{fig:2} for flops and intops. 
A preprocessing stage was applied to each lattice, replacing the generator matrix with another generator matrix for the same lattice via the so-called Lenstra--Lenstra--Lov\'asz (LLL) reduction \cite{ref:7,ref:10}. The operations needed for the reduction were not counted, since the preprocessing is only done once for each lattice, regardless of the number of received vectors. The gain with new algorithms increases linearly with dimension, while the reduction does not change the ratios substantially. The drawback is a somewhat larger number of intops, but the penalty converges to a mere 15$\%$ increase at high dimensions. In simulations it was observed that most of the operations in the algorithms are flops, especially as the dimension increases. For instance, at dimension 60 with the old $\bH$-based algorithm, the flops are roughly 10 times more than the intops. Hence, flops dominate the complexity of the algorithms and intops have a relatively small effect on the overall complexity.

We also measured the running time for the algorithms. As expected, the gain increases roughly linearly with the dimension, similarly to the flops curves in Fig.~\ref{fig:2}. However, the slope of the curve varies significantly between different processors and compilers, which is why we did not include running time in Fig.~\ref{fig:2}. At dimension 60, the gain ranged from 1.7 (AMD processor, Visual C++ compiler) to 2.7 (Intel processor, GCC compiler), for the $\bH$-based algorithm without reduction. We can thus safely conclude that the reduced number of operations translates into a substantial speed gain, but how much depends on the computer architecture.

\subsection{Finite Constellations} \label{sec:5.2}

The channel model in \eqref{eq:16} for an $L$-PAM constellation is considered, where the average symbol energy of the constellation, $E_{s}$, is calculated from the signal set $\{-\frac{L-1}{2},-\frac{L-1}{2}+1,\ldots,\frac{L-1}{2}\}$ and the SNR is defined as $E_{b}/N_{0}$, where $E_{b}=E_{s}/\log_{2}L$ is the average energy per bit and $N_{0}/2$ is the double-sided noise spectral density.
\begin{figure}
\psfrag{Dimension}[cc][][0.8]{Dimension}
\psfrag{Ratio}[cc][][0.8]{Gain}
\psfrag{Flops}[cc][][0.8]{}
\psfrag{G Based SNR=0}[cc][][0.675]{\hspace{14.15mm}$\bG$-based SNR=0dB}
\psfrag{G Based SNR=5}[cc][][0.675]{\hspace{14.1mm}$\bG$-based SNR=5dB}
\psfrag{G Based SNR=10}[cc][][0.675]{\hspace{14.65mm}$\bG$-based SNR=10dB}
\psfrag{H Based SNR=0}[cc][][0.675]{\hspace{13.9mm}$\bH$-based SNR=0dB}
\psfrag{H Based SNR=5}[cc][][0.675]{\hspace{13.75mm}$\bH$-based SNR=5dB}
\psfrag{H Based SNR=10}[cc][][0.675]{\hspace{14.4mm}$\bH$-based SNR=10dB}
\psfrag{0}[cc][][0.8]{0}
\psfrag{10}[cc][][0.8]{10}
\psfrag{20}[cc][][0.8]{20}
\psfrag{30}[cc][][0.8]{30}
\psfrag{40}[cc][][0.8]{40}
\psfrag{50}[cc][][0.8]{50}
\psfrag{60}[cc][][0.8]{60}
\psfrag{70}[cc][][0.8]{70}
\psfrag{80}[cc][][0.8]{80}
\psfrag{1}[cc][][0.8]{1~}
\psfrag{1.5}[cc][][0.8]{1.5~}
\psfrag{2}[cc][][0.8]{2~}
\psfrag{2.5}[cc][][0.8]{2.5~}
\psfrag{3}[cc][][0.8]{3~}
\psfrag{3.5}[cc][][0.8]{3.5~}
\psfrag{4}[cc][][0.8]{4~}
\psfrag{4.5}[cc][][0.8]{4.5~}
\psfrag{5}[cc][][0.8]{5~}
\centering
{\includegraphics[height=5.5cm, width=\columnwidth,
trim=10mm 0mm 35mm 15mm, clip]{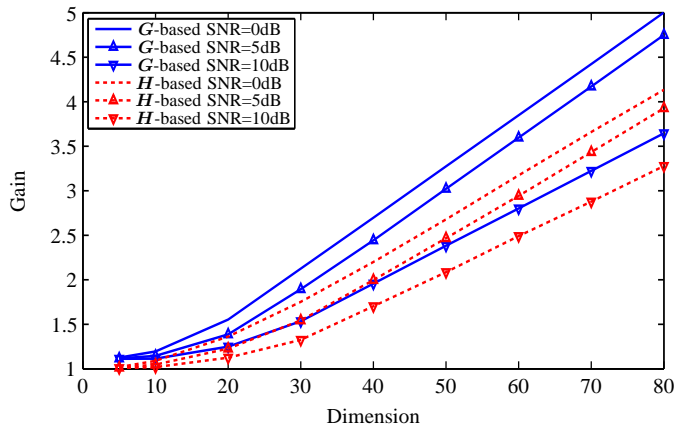}}
\caption{Average gain in the number of flops with the new algorithms for a 2-PAM constellation and various SNRs.} \label{fig:5}
\end{figure}
\begin{figure}
\psfrag{Dimension}[cc][][0.8]{Dimension}
\psfrag{Ratio}[cc][][0.8]{Gain}
\psfrag{Flops}[cc][][0.8]{}
\psfrag{G Based SNR=0}[cc][][0.675]{\hspace{14.15mm}$\bG$-based SNR=0dB}
\psfrag{G Based SNR=5}[cc][][0.675]{\hspace{14.1mm}$\bG$-based SNR=5dB}
\psfrag{G Based SNR=10}[cc][][0.675]{\hspace{14.65mm}$\bG$-based SNR=10dB}
\psfrag{H Based SNR=0}[cc][][0.675]{\hspace{13.9mm}$\bH$-based SNR=0dB}
\psfrag{H Based SNR=5}[cc][][0.675]{\hspace{13.75mm}$\bH$-based SNR=5dB}
\psfrag{H Based SNR=10}[cc][][0.675]{\hspace{14.4mm}$\bH$-based SNR=10dB}
\psfrag{0}[cc][][0.8]{0}
\psfrag{5}[cc][][0.8]{5}
\psfrag{10}[cc][][0.8]{10}
\psfrag{15}[cc][][0.8]{15}
\psfrag{20}[cc][][0.8]{20}
\psfrag{25}[cc][][0.8]{25}
\psfrag{30}[cc][][0.8]{30}
\psfrag{35}[cc][][0.8]{35}
\psfrag{40}[cc][][0.8]{40}
\psfrag{45}[cc][][0.8]{45}
\psfrag{50}[cc][][0.8]{50}
\psfrag{1}[cc][][0.8]{1~}
\psfrag{1.5}[cc][][0.8]{1.5~}
\psfrag{2}[cc][][0.8]{2~}
\psfrag{2.5}[cc][][0.8]{2.5~}
\psfrag{3}[cc][][0.8]{3~}
\psfrag{3.5}[cc][][0.8]{3.5~}
\psfrag{4}[cc][][0.8]{4~}
\psfrag{4.5}[cc][][0.8]{4.5~}
\centering
{\includegraphics[height=5.5cm, width=\columnwidth,
trim=10mm 0mm 35mm 15mm, clip]{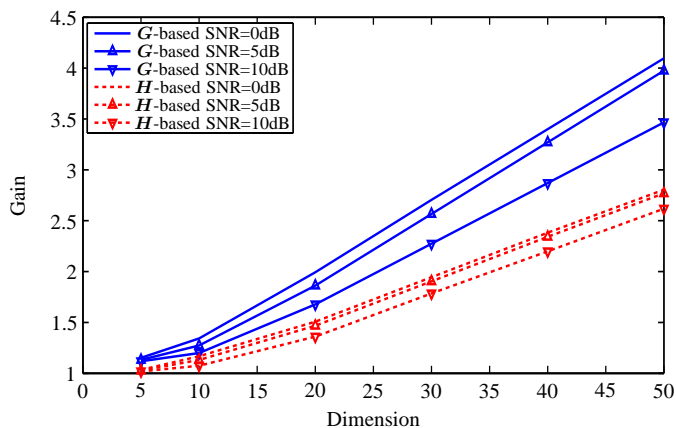}}
\caption{Average gain in the number of flops for a 4-PAM constellation and various SNRs.} \label{fig:6}
\end{figure}

The gain in flops is presented in Figs.~\ref{fig:5}--\ref{fig:6} for 2-PAM and 4-PAM constellations, resp., averaged over 100 random channel matrices $\bG$ with i.i.d.\ zero-mean, unit-variance elements. The same general conclusion as for lattices holds for finite constellations too: The new algorithms provide a substantial complexity gain, and the gain increases linearly with the dimension. However, in contrast to lattice decoding, the gains are here higher for $\bG$-based implementations. Furthermore, the gains increase at low SNR, and 4-PAM offers slightly higher gains than 2-PAM.

\end{document}